\documentclass[aps,prl,twocolumn,showpacs,amsmath,amssymb]{revtex4}

\usepackage{graphicx}
\usepackage{dcolumn}
\usepackage{bm}

\begin{document}
\bibliographystyle{naturemag}
\title{Near-threshold high-order harmonic spectroscopy with aligned molecules}

\author{H. Soifer$^1$, P. Botheron$^2$, D. Shafir$^1$, A. Diner$^1$, O. Raz$^1$, B. D. Bruner$^1$, Y. Mairesse$^2$,
B. Pons$^2$ and N. Dudovich$^1$}

\affiliation{$^1$Department of Physics of Complex Systems, Weizmann
Institute of Science, 76100 Rehovot, Israel\\
$^2$CELIA, Universit\'e de Bordeaux I-CNRS-CEA, 351 Cours de la Lib\'eration, 33405 Talence, France.}

\date{\today}

\begin{abstract}
We study high-order harmonic generation in aligned molecules close
to the ionization threshold. Two distinct contributions to the
harmonic signal are observed, which show very different responses to
molecular alignment and ellipticity of the driving field. We perform
a classical electron trajectory analysis, taking into account the
significant influence of the Coulomb potential on the
strong-field-driven electron dynamics. The two contributions are
related to primary ionization and excitation processes, offering a
deeper understanding of the origin of high harmonics near the
ionization threshold. This work shows that high harmonic
spectroscopy can be extended to the near-threshold spectral range,
which is in general spectroscopically rich.

\end{abstract}

\maketitle

High-harmonic generation (HHG) in strong-field light-matter
interactions is generally understood in terms of a three-step
process: an electron is removed from the target, accelerated, and
driven back to its parent ion by the laser field. Finally, it
recombines into the bound state, and emits a high-harmonic photon
\cite{Cor93}. This process has been extensively studied for
harmonics with energies higher than the ionization threshold
($I_p$), corresponding to frequencies in the extreme ultraviolet
(XUV) range. In this letter, we focus on the intricate region close
to the threshold and use aligned molecules to elicit the dynamics of
near-threshold HHG.

The spectral region near the ionization threshold is, in general,
spectroscopically very rich. In this respect, using near-threshold
harmonic spectroscopy to probe specific structures such as
resonances is an appealing alternative to standard photoionization
spectroscopy. It could help uncover nontrivial attosecond dynamics
occuring near the threshold, such as excitation of electronic
wavepackets or ionization via resonant states
\cite{Haessler09,Mauritsson}. However, the extension of high
harmonic spectroscopy to the threshold region requires a detailed
knowledge of the underlying HHG mechanisms in this specific regime.

Furthermore, high harmonic spectroscopy is being used in a growing
number of experiments to measure the structure and dynamics of small
atomic and molecular systems with unprecedented sub-femtosecond and
Angstr\"om resolution
\cite{Itatani04,Marangos08,Shafir09,Olga09_Nature,McFarland,Haessler10}.
As we move to more complex molecular structures, several orbitals
contribute to the laser-target interaction and nontrivial hole
dynamics are induced. Due to the contribution of multiple orbitals,
there is no clear distinction between near- and above-threshold
harmonics. The ability to identify the different processes in this
regime is an essential step towards decoupling and identifying the
contributions from the various orbitals.

Above the ionization threshold, the intuitive three-step model is
qualitatively confirmed and quantitatively reinforced by the Strong
Field Approximation (SFA)\cite{Lewen94}.  In this model, a photon at
frequency $\omega$ is emitted in the energy range $I_p \leq
\hbar\omega \lesssim I_p + 3.17\cdot U_p$, where $I_p$ is the
ionization potential and $U_p$ the ponderomotive energy.  In the
SFA, the free electron is controlled by the strong laser, whereas
the Coulomb potential is neglected. The SFA predicts many of the
fundamental characteristics of HHG, in particular, the fact that
each harmonic mainly stems from two electronic pathways, commonly
referred to as 
``short" and ``long" trajectories \cite{lhul96}. However, the SFA is
not accurate as we approach the ionization threshold, where the
Coulomb force cannot be neglected \cite{smirnov08}.

There has recently been increased interest in the near-threshold
regime \cite{Yost09,Power10,hostetter10}. Yost \textit{et al.} have
considered the generation of harmonics below the ionization
threshold in xenon atoms \cite{Yost09}. In conjunction with
measurements of harmonic yields as functions of the laser intensity,
they solved the time-dependent Schr\"odinger equation and identified
two distinct contributions to below-threshold harmonics, with
different intensity-dependent phases. The contribution with the
highest intensity dependence of the phase was attributed to long
electron trajectories, whereas the second contribution, with smaller
intensity dependence, was related to primary multiphoton processes.
Hostetter {\em et al.} \cite{hostetter10} supported this observation
from a semiclassical point of view.

How can we directly and experimentally distinguish between the
different HHG processes in the vicinity of the ionization threshold?
There are two important knobs that can be used to manipulate the HHG
process. The first knob is related to the structure of the bound
state and can be manipulated by producing harmonics from aligned
molecules. The second is related to the free electron dynamics and
can be tuned by controlling the laser ellipticity. Here we combine
the two knobs, using extensively studied N$_2$ and O$_2$ molecules,
and manipulate both the target structure, through alignment, and the
laser field, which guides the free electron wavepacket. Such a
manipulation is applied to elicit the dynamical features of
threshold HHG. In addition, we perform a theoretical analysis in
terms of classical electron trajectories, which yields an intuitive
picture of the interaction and improves our understanding of the
underlying mechanisms.

The experiment is composed of two parts. We control the molecules
using a pump-probe setup in which the pump creates rotational
wavepackets and the probe generates the harmonic signal
\cite{Itatani05}. Once the molecular axis is fixed we manipulate the
free electron's dynamics by scanning the ellipticity of the probe
beam. High harmonics are generated with an 800-nm, 30-fs, 1-KHz
laser beam, focused into a pulsed gas jet at $\sim 1.5\times10^{14}
$ W/cm$^2$ intensity. As a pump we use part of the 800-nm beam,
which we focus at a few 10$^{13}$ W/cm$^2$ in the jet. The probe
ellipticity $\epsilon$ is controlled by rotating a half waveplate in
front of a fixed quarter waveplate in order to keep the main axis of
the polarization ellipse fixed. $\epsilon$ is defined through
$F(t)=F_0\cos(\omega t)\hat{x}+i\epsilon F_0\sin(\omega t)\hat{y}$
and the total intensity $|F|^2$ is kept constant throughout the
experiment. We perform this scan for alignment angles of 0 and 90
degrees. The harmonics are dispersed by a 1200 mm$^{-1}$ grating and
imaged on microchannel plates. The cylindrical geometry of the
grating, combined with an optimized adjustment of the position of
the gas jet with respect to the laser focus, enables us to
discriminate between short and long trajectories, spectrally and
spatially, in both below- and above-threshold regimes
\cite{Salieres05,Yost09}. In the following, we will focus on the
spectral range in the vicinity of the ionization potentials
of N$_2$ and O$_2$, which respectively correspond to 10.05
$\omega_0$ and 8.78 $\omega_0$, where $\omega_0$ is the fundamental
laser frequency.

\begin{figure}
\includegraphics[width=0.43\textwidth,clip]{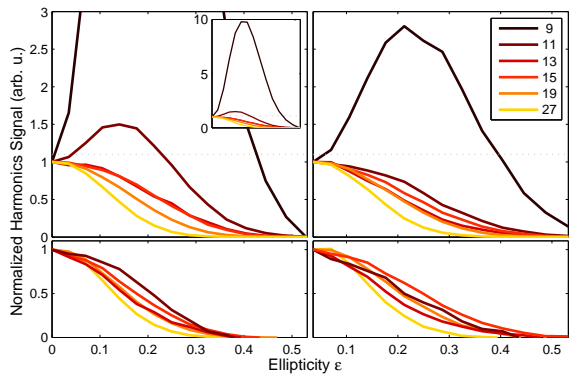}
\caption{Harmonic yields in O$_2$ aligned at 0 (left) and 90 (right)
degrees, for long (bottom) and short (top) trajectories, as
functions of the ellipticity $\epsilon$ of the probe pulse. The
signal from each harmonic is normalized to its value at
$\epsilon=0$. The inset shows the full scale of the increase in
harmonic 9.}
\end{figure}

\begin{figure}
\includegraphics[width=0.43\textwidth,clip]{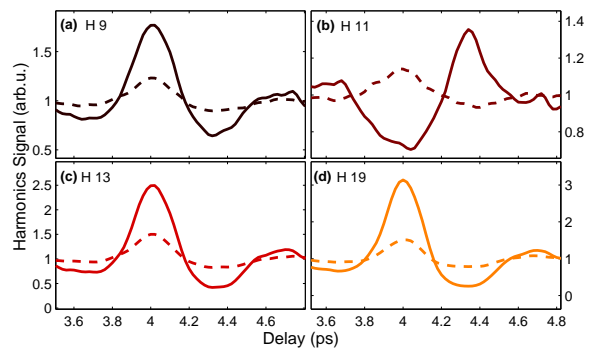}
\caption{Normalized harmonic yields in rotationally excited N$_2$
for harmonics 9-13,19, as a function of the pump-probe delay. The
solid and dashed lines correspond to the short and long
trajectories, respectively.}
\end{figure}

Figure 1 shows the harmonic signal as a function of the laser
ellipticity for O$_2$ molecules for alignment angles of 0 (left) and
90 (right) degrees. Long trajectories were measured down to harmonic
11. As we increase the laser ellipticity, all the long trajectories'
signals (bottom panels) decrease significantly. Such a response has
been explained by the SFA for above-threshold harmonics
\cite{Antoine96}: the ellipticity of the driving field induces a
lateral shift of the free electron wavepacket, and therefore reduces
the recombination probability. In the vicinity of the ionization
threshold, we observe a similar effect.

As we switch from long to short trajectories, a striking difference
appears. For an alignment angle of 0 degrees, harmonics 9 and 11
increase as a function of the ellipticity and reach a maximum at
$\epsilon\sim0.2$ and 0.15, respectively. The signal at this point
is enhanced by a factor of $9.8\pm0.7$ for the $9^{th}$ harmonic and
by $1.5\pm0.1$ for the $11^{th}$ harmonic. A similar effect is
observed for an alignment angle of 90 degrees, but only for the
$9^{th}$ harmonic ($2.7\pm0.2$). When the intensity $I$ is slightly
increased (from $\sim1.6$ to $\sim1.8\times10^{14}$ W/cm$^2$) this
enhancement is observed in an additional harmonic - 13 and 11 for 0
and 90 degrees, respectively. Well above the ionization threshold,
all short trajectories responses exhibit the usual decaying rate,
regardless of the intensity.

Repeating the experiment with N$_2$ molecules, no anomalous
responses are observed. All harmonic orders decay smoothly as a
function of $\epsilon$, exhibiting similar decay rates for both
short and long trajectories (not shown). So, while all long
trajectories follow the expected ellipticity response as predicted
by the intuitive three-step model, the reactions of near-threshold
short trajectories are system-dependent and highly selective. The
observed enhancements at $\epsilon \neq 0$ cannot be related to the
structure of the ground molecular orbital, and related interference
processes at the time of recombination \cite{Kanai07}, since these
latter would affect both short and long trajectories. We shall
return to this point later on, in the context of our theoretical
analysis.

An additional difference between short and long trajectories in
N$_2$ molecules near $I_p$ is observed when the pump-probe delay is
scanned using linearly polarized beams. Figure 2 shows the results
of our pump-probe scans around the half-revival of N$_2$
\cite{Itatani05}, for both short and long trajectories. As observed
in previous experiments, the total harmonic signal peaks at a delay
of 4.1 ps, when the molecules are aligned along the polarization
axis of the pump pulse, and has a minimum at 4.3 ps, when the
molecules are anti-aligned (see Fig. 2 a,c, and d for harmonics 9,
13, and 19). Long trajectories were measured down to the $9^{th}$
harmonic, with the typical revival structure (dashed lines in Fig.
2). However, the $11^{th}$ harmonic has surprising features. Whereas
the long trajectory exhibits the typical revival shape, with
contrast $\frac{max-min}{max+min}=0.10\pm0.03$, the short one (solid
line) displays an inverted structure
($\frac{max-min}{max+min}=0.35\pm0.04$). In order to verify that the
inversion effect we observe does not depend on propagation, we
performed a systematic study by varying the gas pressure by one
order of magnitude and found similar results. An intensity scan from
1 to 2$\times10^{14}$ W/cm$^2$ also shows the robustness of this
effect. Only below 1.2$\times10^{14}$ W/cm$^2$ does the inversion of
harmonic 11 disappear.

Both experiments - the ellipticity and revival scans - indicate that
threshold harmonics are generated by two completely different
processes. Whereas the long trajectories behave as higher harmonic
orders and therefore seem to be generated according to the standard
three-step model, it is clear that the short trajectories originate
from a different mechanism. How can threshold harmonics be explained
in the framework of classical trajectories? What breaks the symmetry
between the two families of electron trajectories?

In order to answer these fundamental questions, we have performed
statistical Classical Trajectory Monte Carlo (CTMC) calculations
\cite{Botheron09} that treat the laser $V_F$ and Coulomb $V_C$
potentials on the same footing. We consider the prototypical N$_2$
system described in terms of a one-dimensional scaled hydrogen atom
with an effective nuclear charge $Z_{e}$ such that $Z_e=\sqrt{2\
I_p}$. The CTMC procedure employs an N-point discrete representation
of the phase-space distribution,
$\varrho(x,p,t)=\frac{1}{N}\sum_{j=1}^{N}{\delta(x-x_j(t))\delta(p-p_j(t))}$,
in terms of N non-interacting trajectories
$\{x_j(t),p_j(t)\}_{j=1,..,N}$. The initial condition,
$\varrho(x,p,t=0)$, consists of the Wigner phase-space distribution
associated with the fundamental classical energy bin, as detailed in
\cite{Botheron09}.

\begin{figure}
\includegraphics[width=0.42\textwidth,clip]{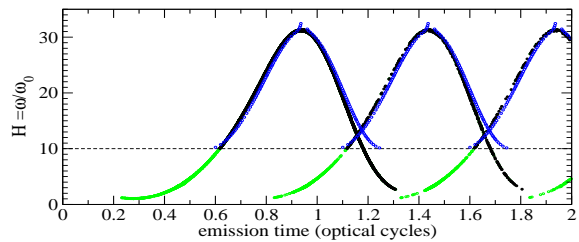}
\caption{Calculated harmonic order as a function of time of emission
$t'$, for N$_2$ embedded in a two-cycle field with $\lambda=800$ nm
and $I=1.5\ 10^{14}$ W/cm$^2$. The blue (empty) circles correspond
to SFA calculations, whereas black (dark) and green (light) dots,
respectively, represent harmonics stemming from ionizing and
excitation trajectories, in the framework of CTMC-Wigner
calculations.}
\end{figure}

\begin{figure}
\includegraphics[width=0.43\textwidth,clip]{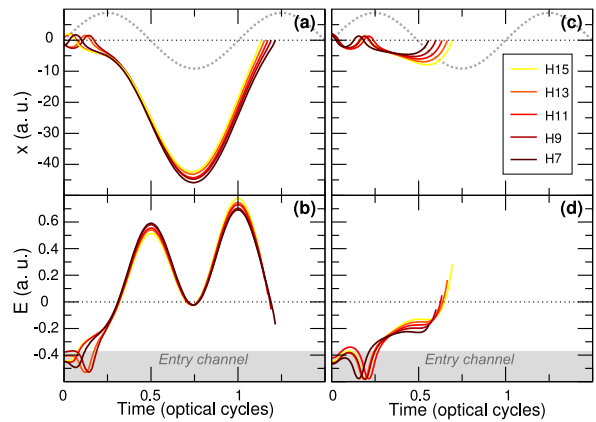}
\caption{Calculated electron position $x$ (a,c), and energy $E$
(b,d) as functions of time $t$, for some short (right column) and
long (left column) trajectories of Fig. 3. The dotted line in (a,c)
represents the evolution of the electric field $F(t)$.}
\end{figure}

The CTMC picture of HHG is obtained by considering an instantaneous
recombination to the ground state at time $t'$ whenever a trajectory
with energy $E(t')=p(t')^2/2+V_C(t')$ returns to $x(t')=0$ after
having been ionized at time $t''$. Such a process is accompanied by
the emission of a photon with energy $E(t')+I_p$. Figure 3 presents
the harmonic order of the photon emitted upon recombination as a
function of time $t'$ of emission. Our CTMC results are compared to
their SFA counterparts. For above-threshold emission, CTMC and SFA
results are in good agreement. In contrast to the SFA, the CTMC
method yields long trajectories for below-threshold harmonics; in
the ($t'$,H) plane of Fig. 3, these trajectories appear as
continuations of the above-threhold ones below $I_p$.

To illustrate how these trajectories succeed in returning to the ion
with $E(t')<0$, we present in Fig. 4 the temporal evolution of the
electron position $x(t)$ and of the energy $E(t)=p(t)^2/2+V_C(t)$,
for some typical long trajectories. Ionization occurs at $t'' \sim
0.3\ T_0$, where $T_0$ is the laser period, soon after the field
maximum. At this time, the dispersion of the free wavepacket in
coordinate space, although small, is sufficient to allow the
subsequent differentiation of the trajectories in the combined laser
and Coulomb fields. The electron returns to the origin $x=0$ at
$0.95 T_0 \lesssim t'' \lesssim 1.3 T_0$. Earlier electrons emit
higher harmonics. At the time of recollision, the laser force
$-F(t)$ acting on the electron is in the opposite direction to the
Coulomb force \cite{Yost09}. Negative energy $E(t')$ is enabled for
those later trajectories that recombine at $t' \sim 1.25 T_0$, where
$p^2(t')/2 + V_C(t')<0$. Our CTMC model thus shows clear continuity
between above- and below-threshold harmonics for the long
trajectories. All of them obey the standard three-step process and
therefore have similar responses to ellipticity and revival scans.

When the same analysis is applied to short trajectories, no
below-threshold harmonics appear (see Fig. 3). We thus modify our
model by relaxing the condition regarding ionization: we only
require the trajectory to leave the fundamental classical energy bin
before it returns back to $x=0$, but we do not require it to be
ionized (see Fig. 4). In other words, we allow excitation pathways
to contribute to HHG. The results corresponding to this new
criterion are included in Fig. 3; below-threshold short trajectories
emerge, which is consistent with our experimental observations. They
appear as the continuation below $I_p$ of the above-threshold short
trajectories. Their temporal evolution is illustrated in Fig.
4(c-d). The short trajectories corresponding to $\omega \gtrsim I_p$
are ionized shortly before recombination, whereas for
below-threshold HHG, the trajectories recombine before ionization,
with $E(t')<0$. Inversely to what occurs for long trajectories,
earlier electrons emit lower harmonics. Clearly, part of the
below-threshold HHG signal stems from excitation processes, which
are classically described in terms of short trajectories.
Interestingly, a Bohmian description of the interaction yields
quantum trajectories that nicely match, in both position and energy
scales, the short CTMC trajectories \cite{Botheron10}. Those
trajectories do not involve tunnel-ionization, in agreement with the
semiclassical calculations of \cite{hostetter10}, and therefore are
beyond the scope of the conventional three-step model.

We can now reinterpret the experimental observations concerning
short trajectory below-threshold harmonics in light of their
excitation origin. With respect to the ellipticity dependence
observed in Fig. 1, a description of excitation in terms of a
non-resonant absorption of H photons leads to harmonic yields that
smoothly decrease as $\epsilon$ increases, in agreement with the
N$_2$ shapes. This regular behavior is modified if intermediate
resonances come into play during the multiphoton absorption, as
shown in \cite{miyazaki} for atomic targets. As $\epsilon$
increases, the interplay between the decreasing linear and
increasing circular field components then yields
$\epsilon$-responses similar to those of $H=9-13$ in O$_2$.
Obviously, the occurrence of resonances depends not only on the
target considered, but also on the alignment angle that determines
the allowed symmetry of the intermediate excited states. Resonantly
enhanced excitation may also induce the inversion of revival
structures. Depending on the symmetries of the intermediate states
involved in the multiphotonic pathway, the harmonic strength can be
maximual for a different molecular alignment than in the
conventional three-step process involving tunneling. Finally, it is
well known in strong-field physics that the ionization threshold of
a target embedded in an intense laser field is subject to an
ac-Stark ascending shift proportional to $F^2$. In this respect, the
short trajectories associated with harmonic orders close to the
unperturbed $I_p$ recombine at time $t'$, where $F(t') \neq 0$.
Therefore, harmonic orders $H \gtrsim I_p/\omega_0$, which belong to
above-threshold HHG at low $I$, fall down in the below-threshold
regime as $I$ increases. Accordingly, we understand the appearance
of additional anomalous responses to $\epsilon$ in O$_2$ as $I$
increases, and why the inversion of harmonic 11 disappeared in N$_2$
for $I \leq 1.2\times 10^{14}$ W/cm$^2$, where $H=11$ then
effectively corresponds to above-threhold HHG.

To sum up, our joint experimental/theoretical study of harmonic
generation in aligned N$_2$ and O$_2$ has allowed us to observe and
explain, unambiguously, two main contributions to near-threshold
HHG. The long electron trajectories belong to the three-step model,
and encode the structural information in the same way as in
conventional high harmonic spectroscopy. The short trajectories stem
from multiphoton-driven pathways, and can be used to reveal
complementary information on excited states of the molecule. The
ellipticity and alignment responses of the short trajectory signal
will provide information on the symmetry and dynamics of the
intermediate states involved in the multiphoton process. This
approach adds to the ability of HHG spectroscopy to follow a
molecular orbital during chemical dynamics \cite{Worner10}. Since
this method is extremely sensitive to the location of $I_p$, it is
possible to probe changes in the threshold itself during chemical
processes. Thus, our study is an integral part of the ongoing effort
to resolve the structure as well as the dynamics of molecular
orbitals.

\end{document}